\documentclass[dvipdfm,twocolumn,showpacs,preprintnumbers,amsmath,amssymb]{revtex4}

\usepackage{bm}
\usepackage[dvips]{graphicx,color}
\usepackage[mathscr]{eucal}
\usepackage[dvipdfm,colorlinks]{hyperref}
\hypersetup{
    bookmarksopen=true,
    pdfpagemode=UseOutlines,
    pdfstartview=FitH,
    linkcolor=blue,citecolor=red,urlcolor=green}

\allowdisplaybreaks

\begin{document}

\title{Exact Equal Time Statistics of Orszag-McLaughlin Dynamics By The Hopf Characteristic Functional Approach}

\author{Ookie Ma\footnote{e-mail: seungwook\_ma@physics.brown.edu} 
and J. B. Marston\footnote{e-mail: marston@physics.brown.edu}}
\affiliation{Department of Physics, Brown University, Providence,
Rhode Island 02912, USA}

\date{\today}

\begin{abstract}
By employing Hopf's functional method, we find the exact characteristic functional for a simple nonlinear dynamical system introduced by Orszag.  Steady-state equal-time statistics thus obtained are compared to direct numerical simulation.  The solution is both non-trivial and strongly non-Gaussian.
\end{abstract}


\maketitle

\section{Introduction}
\label{intro}

A nonlinear dynamical system can be completely deterministic and its solution unique, yet two trajectories that begin close to one another may diverge significantly in a finite time.  Such sensitive dependence on initial conditions sets a fundamental limit on predictive accuracy, as these systems forget their initial conditions after a short time.  Probabilistic descriptions, by contrast, avoid the details of time-evolution, and instead answer meaningful statistical questions.  As a canonical example, experimental measurements of turbulent velocity fluctuations show disordered and unpredictable behavior yet reproducible statistical properties.  The complexity of the flow contrasts with the smoothness of averaged quantities like the energy spectrum, demonstrating that statistical descriptions may be both economical and insightful.

To accumulate statistics, numerical simulations may implement ensemble averages over initial condition and/or long-time integration, yet the computational work required to carry out such calculations can be prohibitive in the case of high-dimensional systems.  Purely statistical approaches can be quite helpful for such problems.  Furthermore, sub-grid physics is often modeled statistically; fewer inconsistencies will arise if all of the dynamical variables are treated statistically from the outset.  Finally, from a theoretical perspective it can be more enlightening to work directly with direct statistical approaches \cite{lorenz}.  

In this paper we apply a method pioneered by Hopf \cite{hopf} to determine the exact characteristic functional of a simple dynamical system introduced by Orszag \cite{orszaglec}.  Knowledge of the characteristic functional yields all of the equal-time moments and offers some advantage over the more common statistical description in terms of the probability density function (PDF).  In particular, moment and cumulant expansions arise naturally in the Hopf formalism \cite{frisch}.  Such expansions, however, suffer from the usual closure problem and break down when the statistics are strongly non-Gaussian.  The exact solution to Hopf's equation is both non-trivial and, for a finite number of dimensions, non-Gaussian.  Exact solutions are invaluable as they both yield physical insight and provide important benchmarks against which approximate methods can be tested.

The outline of the paper is as follows.  In Section~\ref{model} we introduce the finite-dimensional Orszag-McLaughlin dynamical system.  We briefly review the Hopf functional approach in Section~\ref{hopf}.  The exact characteristic functional of the Orszag-McLaughlin system is presented in Section~\ref{exact}.  In Section~\ref{infinite} we consider the limit of large dimension and show that the statistics become Gaussian in this limit.  Comparison of the solution against direct numerical simulation is made in Section~\ref{numerics}.  Finally in Section~\ref{discussion} we highlight a few points.

\section{Orszag-McLaughlin Dynamical System}
\label{model}

\begin{figure}
\centerline{\includegraphics[scale=0.52]{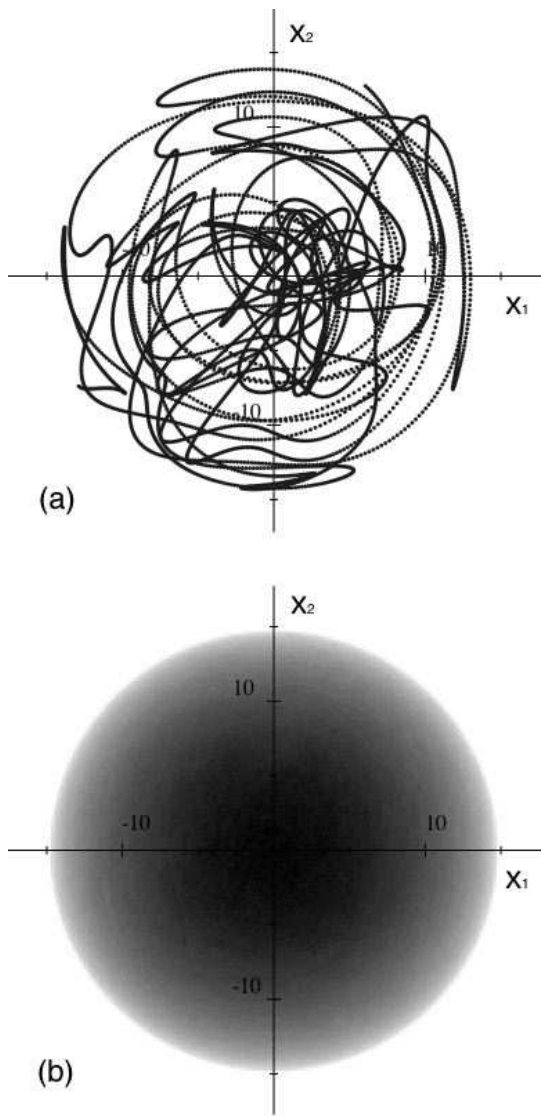}}
\caption{(a) A typical trajectory of the Orszag-McLaughlin dynamical system, for the case $N = 5$, projected onto the $x_1$-$x_2$ plane.  Trajectories appear random and lie on the surface of a 5-dimensional hypersphere.  (b) The probability density function projected onto the $x_1$-$x_2$ plane, as determined by long-time integration.}
\label{trajectory}
\end{figure}

In Orszag's lecture notes \cite{orszaglec} and in a subsequent paper with McLaughlin \cite{orszmcla}, a finite-mode system was introduced as a toy model of inviscid flow:
\begin{eqnarray}
\frac{dx_i}{dt} = x_{i+1}x_{i+2}+x_{i-1}x_{i-2}-2x_{i+1}x_{i-1},
\label{eom}
\end{eqnarray}
with $ i = 1, \ldots, N$ and periodic boundary condition $x_{i+N} = x_i$.  The dimension $N$ can be either even or odd; we focus on odd $N$ here.
Like Euler's equation, the equations of motion (EOM) contain only quadratic terms.  Though the EOM are simple, they generate complicated trajectories as shown in Fig.~\ref{trajectory}.

Despite the fact that the Orszag-McLaughlin dynamical system  is non-Hamiltonian (trivially so in an odd number of dimensions), it conserves an energy-like quantity in the sense that
\begin{equation}
E=\frac{1}{2}\sum_{i=1}^N x_i^2
\label{energy}
\end{equation}
is a constant of the motion.  Trajectories lie on a hypersphere of radius:
\begin{equation}
R=\sqrt{2E}\ .
\label{radius}
\end{equation}
For odd values of $N\geq5$, Eq.~(\ref{radius}) is the only isolating integral of motion \cite{orszmcla}.
In addition, for $N=5$, all periodic orbits and fixed points are unstable \cite{orszmcla}.  The system appears to be ergodic for all initial conditions except a set of measure zero.  This also seems to be the case for other odd values of $N>5$.  Finally we note that the dynamical system formally obeys Liouville's Theorem, despite being non-Hamiltonian:
\begin{eqnarray}
0 &=& \sum_{i=1}^N\frac{\partial}{\partial x_i} \left( \frac{dx_i}{dt}\right)
\nonumber \\
&=& \sum_{i=1}^N\frac{\partial}{\partial x_i}(x_{i+1}x_{i+2}+x_{i-1}x_{i-2}-2x_{i+1}x_{i-1})\ .
\label{liouville}
\end{eqnarray}

The assumption of ergodicity combined with the energy constraint Eq.~(\ref{energy}) has an important consequence.   Strikingly, the system exhibits continuous rotational symmetry in its statistics, despite the fact that the EOM are invariant only under discrete rotations corresponding to the permutation of indices.  The statistics are determined by only a single parameter, the energy $E$, or equivalently the radius $R$.

\section{Hopf Functional Approach}
\label{hopf}

Hopf developed a functional method to access the equal-time statistics of deterministic dynamical systems.  In Ref.~\onlinecite{hopf}, Hopf derived but did not attempt to solve an equation governing the characteristic functional for the turbulent velocity field of an incompressible flow.  Here we briefly review his approach.  The reader can find more detailed information in Ref.~\onlinecite{frisch}.  

For the purposes of illustrating the method, and applying it in the next Section to the Orszag-McLaughlin system, it suffices to consider deterministic dynamical systems described by a set of first-order differential equations:
\begin{equation}
\frac{dx_i}{dt}=F_i(x_1, x_2, \ldots, x_N);\ \ i = 1, \ldots, N.
\label{firstorder}
\end{equation}
Introduce a set of variables $\vec u$ conjugate to the coordinates $\vec x$ and a functional $\Psi$ that resembles a wavefunction in quantum mechanics:
\begin{equation}
\Psi[\vec{u}, \vec{x}(t)] \equiv e^{i \vec u \cdot \vec x(t)}\ .
\label{unavpsi}
\end{equation}
It then follows from the EOM, Eq.~(\ref{firstorder}), that $\Psi$ is governed by a Schr$\rm{\ddot{o}}$dinger-like equation:
\begin{equation}
i\frac{d \Psi}{dt}=\hat H \Psi\ .
\label{hopfeq}
\end{equation}
Here the Hopf ``Hamiltonian" on the right-hand side is the linear operator:
\begin{eqnarray}
\hat H = -\sum_{j=1}^N u_j F_j\left( \frac{1}{i}\frac{\partial}{\partial u_1}, 
\frac{1}{i}\frac{\partial}{\partial u_2}, \ldots, \frac{1}{i}\frac{\partial}{\partial u_N} \right)\ .
\label{hamiltonian}
\end{eqnarray}
Note that  $\hat H$ is generally non-Hermitian, as there is no requirement (unlike in quantum mechanics) that the time-evolution of $\Psi$ be unitary.  For instance the inner product, which plays a central role in quantum mechanics, has no particular meaning in the Hopf formalism.  We call Eq.~(\ref{hopfeq}) together with Eq.~(\ref{hamiltonian}) ``Hopf's equation'' in the following.

We emphasize that Hopf's equation is linear even though the EOM may be nonlinear.  Therefore -- and this is the key point -- Eq.~(\ref{hopfeq}) can be averaged over an ensemble of initial conditions.  It is the averaging that blurs the deterministic trajectories and, pursuing the analogy with quantum mechanics, can be thought of as the origin of non-zero $\hbar$. The corresponding characteristic functional $\overline{\Psi}(\vec{u}, t)$ that solves the Hopf equation is simply an average of $\Psi[\vec{u}, \vec{x}(t)]$ over initial conditions at time $t=0$:
\begin{eqnarray}
\overline{\Psi}(\vec{u}, t) &=& \overline{e^{i \vec u \cdot \vec x(t)}}
\nonumber \\
&=& \int_{-\infty}^\infty e^{i \vec u \cdot \vec x(t)} P(\vec x(0)) d^N x(0)\ .
\label{psi}
\end{eqnarray}
For notational simplicity we eliminate the bar over $\overline{\Psi}$ in the following.

It is readily seen that derivatives of the characteristic functional yield the equal-time moments:
\begin{eqnarray}
\frac{1}{i}\frac{\partial\Psi(\vec{u}, t)}{\partial u_j}\bigg\vert_{\vec u=0}&=&\int_{-\infty}^{\infty}x_j(t) P(\vec{x}(0)) d^N x(0)\nonumber \\ &=& \overline{x_j(t)}\label{firstmoment},\\
-\frac{\partial^2\Psi(\vec{u}, t)}{\partial u_j \partial u_k}\bigg\vert_{\vec u=0}&=&\int_{-\infty}^{\infty}x_j(t) x_k(t) P(\vec {x}(0)) d^N x(0)\nonumber \\ &=& \overline{x_j(t) x_k(t)},\label{secondmoment}
\end{eqnarray}
and so on.  All conjugate coordinates $\vec{u}$ are taken to zero after derivatives are taken.   From Eq.~(\ref{psi}) it is clear that the characteristic functional is simply the Fourier transform of the PDF, and one uniquely determines the other.  We note that the calculation of n-point correlation functions are more simply extracted from the characteristic functional than from the PDF.  In the case of the PDF we must integrate over all of the coordinates.  For the characteristic functional all that is required is the calculation of some derivatives followed by setting all of the conjugate coordinates to zero \cite{monyag}.  The advantage is particularly clear in the case of infinite-dimensional problems such as fluid flow for which the velocity field is a continuous function of position.

Of special importance are zero-mode characteristic functionals that satisfy $\hat H \Psi_0 = 0$.  The zero-mode plays a role as important as the ground state in quantum mechanics as it encodes information about the steady-state behavior of the dynamical system.  In the following, we employ Dirac's bracket notation $\langle \ldots \rangle$ to indicate steady state statistical averages.

In order for $\Psi$ to be a valid characteristic functional, it must obey two general boundary conditions that follow directly from Eq.~(\ref{psi}):
\begin{eqnarray}
\Psi(\vec 0)&=&1 
\label{bc1}
\end{eqnarray}
and
\begin{eqnarray}
\lim_{|\vec u|\rightarrow\infty}\Psi(\vec u)&=&0\ .
\label{bc2}
\end{eqnarray}
Furthermore, a necessary and sufficient condition to ensure semi-positivity of the PDF is specified by Bochner's Theorem \cite{bochner}:
\begin{eqnarray}
\int d^N u~ d^N u^\prime~ \phi(\vec{u}) \Psi(\vec{u} - \vec{u^\prime}) \phi^*(\vec{u^\prime}) \geq 0
\label{bochner}
\end{eqnarray}
for all complex-valued test functions $\phi(\vec{u})$.  This constrains all of the equal-time moments or cumulants.  For example, even cumulants cannot be negative \cite{kraich,hantalk}.  
Additional boundary conditions arise from conservation laws.  In particular, for the Orszag-McLaughlin system, conservation of energy imposes an infinite set of boundary conditions:
\begin{equation}
\left(-\vec\nabla_u^2\right)^n \Psi\bigg\vert_{\vec u=0}=E^n
\label{econs}
\end{equation}
for all integer $n$.

For completeness, we conclude this brief review of the Hopf functional approach by discussing the inclusion of random forcing.  For simplicity, consider the introduction of Gaussian random forcing $f_i$, $\delta$-correlated in time, and with zero mean.  The EOM are modified to read:
\begin{equation}
\frac{dx_i}{dt}=F_i(\vec x)+ f_i(t)
\end{equation}
where
\begin{eqnarray}
\left[ f_j(t) \right] &=& 0, 
\\
\left[ f_j(t) f_k(t^\prime) \right] &=&2\Gamma \delta_{jk} \delta(t-t^\prime)
\end{eqnarray}
and the square brackets $[ \ldots ]$ denote an average over the random variable.
The Hopf equation for the characteristic functional now reads \cite{novikov}:
\begin{equation}
i\frac{\partial \Psi(\vec{u}, t)}{\partial t}=\left( \vec{u} \cdot \vec{F}( -i \vec{\nabla}_u ) - i \Gamma |\vec{u}|^2 \right)
\Psi(\vec u,t)\ .
\label{randomhopf}
\end{equation}

One possible derivation of Eq.~(\ref{randomhopf}) begins with the better-known Fokker-Planck 
equation \cite{risken}:
\begin{equation}
\frac{\partial P(\vec{x}, t)}{\partial t} =
\left(- \vec{\nabla}_x \cdot \vec{F}(\vec x) + \Gamma \nabla^2_x \right) P(\vec x,t)\ .
\label{fokkerplanck}
\end{equation}
The Hopf equation, Eq.~(\ref{randomhopf}), is recovered upon making the duality transformation $x_j \rightarrow \frac{1}{i} \frac{\partial}{\partial u_j}$ and $\frac{\partial}{\partial x_j} \rightarrow \frac{1}{i}u_j$.  The duality transformation respects the commutation relation between the conjugate coordinates: $[x_j,~ \frac{1}{i}u_k] = \delta_{jk}$.  Thus the Hopf and Fokker-Planck approaches are seen to be equivalent, and dual to one another, yet the inclusion of random forcing seems more intuitive in the latter case.  As is well-known, the first term in the Fokker-Planck equation is a statement of conservation of phase-space density.  The second is a diffusion term due to random forcing, which smears out an initially sharp distribution.   Pawula \cite{pawula} showed that Gaussian random forcing leads to only a finite number of terms in the Fokker-Planck equation.  If the forcing is non-Gaussian, the Fokker-Planck equation must include an infinite set of terms involving derivatives higher than two.

\section{Exact Characteristic Functional}
\label{exact}

The Hopf Hamiltonian for the Orszag-McLaughlin dynamical system follows directly from Eqs.~(\ref{eom}) and (\ref{hamiltonian}):
\begin{eqnarray}
\hat H=\sum_{i=1}^N&&u_i\bigg{(}\frac{\partial^2}{\partial u_{i+1}\partial u_{i+2}}+\frac{\partial^2}{\partial u_{i-1}\partial u_{i-2}}\nonumber\\ &&-2\frac{\partial^2}{\partial u_{i+1}\partial u_{i-1}}\bigg{)}
\end{eqnarray}
where periodic boundary conditions $\partial / \partial u_{i+N} = \partial / \partial u_i$ are implied.  For this particular system the Hamiltonian is Hermitian, though (as noted above) this will not always be the case for other dynamical systems.  

It is straightforward to show that any $\Psi(\vec{u})$ that is a function purely of the magnitude of $\vec{u}$ is a zero-mode of the above $\hat{H}$.  This is consistent both with the assumption of ergodicity, and with the numerical simulation of the PDF illustrated in Fig.~\ref{trajectory}.  However, such zero-modes are, in general, a superposition of states of different energies, Eq.~(\ref{econs}).  Imposition of the energy constraint Eq.~(\ref{econs}) picks out the desired solution.  Consider a Taylor series expansion of $\Psi_0$ in the dimensionless parameter $(R |\vec u|)^2$:
\begin{equation}
\Psi_0(\vec{u})= \sum_{\ell=0}^\infty b_\ell (R |\vec u|)^{2 \ell}\ .
\label{taylor}
\end{equation}
The series coefficients can be found by observing that the energy constraint requires
\begin{equation}
R^{2 \ell} = \langle |\vec{x}|^{2 \ell} \rangle\ ,
\end{equation}
and furthermore, the moments may be expressed in terms of the coefficients $b_\ell$.  For example, successive differentiation of Eq.~(\ref{taylor}) reveals that
\begin{equation}
\langle x_1^4 \rangle=4!~ b_2~ R^4,
\end{equation}
and
\begin{equation}
\langle x_1^2 x_2^2 \rangle= 2~ 2!~ 2!~ b_2~ R^4.
\end{equation}
Combining this with the energy constraint
\begin{eqnarray}
R^4&=&N\langle x_1^4 \rangle+2 {\binom{N}{2}} \langle x_1^2 x_2^2 \rangle
\nonumber \\
&=&N 4!\,R^4 b_2+ 2^2 {\binom{N}{2}} \,2!\,2!\,R^4 b_2
\end{eqnarray}
yields
\begin{equation}
b_2=\frac{1}{N 4!+ 2^2 {\binom{N}{2}} \,2!\,2!}.
\end{equation}
Each coefficient can be determined in this manner.  The series so generated is:
\begin{equation}
\Psi_0(\vec{u}) = \Gamma(N/2) \sum_{\ell=0}^\infty \frac{1}{\Gamma(\ell+N/2)} \frac{(-1)^\ell }{2^{2\ell} \ell!}(R |\vec u|)^{2 \ell}\ .
\label{seriespsi}
\end{equation}
The series may be summed into the Bessel function $J_{N/2-1}$:
\begin{eqnarray}
\Psi_0(\vec u) = \Gamma(N/2)~ (R |\vec u| / 2)^{1-N/2}~ J_{N/2 - 1}(R |\vec u|)\ . 
\label{exactpsi}
\end{eqnarray}
We note that the exact zero-mode characteristic functional satisfies the boundary conditions, Eqs.~(\ref{bc1}) and (\ref{bc2}).   

The characteristic zero-mode may be approached from another direction, one that does not rely upon series expansion.  Knowing that the trajectories are distributed uniformly on the hypersphere's surface, the stationary PDF is a simple $\delta$-function shell:
\begin{equation}
P(\vec x) = \frac{1}{S_{N-1}}~ \delta(|\vec x|-R)
\label{shellpdf}
\end{equation}
where $R$ is given by Eq.~(\ref{radius}).  The normalization factor $S_{N-1}^{-1}$ is simply the inverse of the  total surface area of the hypersphere.  By direct substitution it can be seen that this is a zero-mode solution to the Fokker-Planck  Eq.~(\ref{fokkerplanck}).  
Now $\Psi_0$ may be obtained from the Fourier transform of Eq.~(\ref{shellpdf}).  Due to the spherical symmetry it is most convenient to work in spherical coordinates.  The N Cartesian coordinates are related to spherical coordinates by:
\begin{eqnarray}
x_1 &=& r \sin\phi_0\prod_{j=1}^{N-2}\sin\phi_j
\nonumber \\
x_2 &=& r \cos\phi_0\prod_{j=1}^{N-2}\sin\phi_j
\nonumber \\
x_3 &=& r \cos\phi_1\prod_{j=2}^{N-2}\sin\phi_j\ , {\rm etc.}
\end{eqnarray}
In these coordinates the volume element is:
\begin{equation}
dV = \prod_{j=1}^N dx_j = r^{N-1}dr \prod_{j=0}^{N-2} \sin^j \phi_j~ d\phi_j\ . 
\end{equation}
By exploiting spherical symmetry, we obtain the characteristic functional, Eq.~(\ref{exactpsi}):
\begin{eqnarray}
\Psi_0(\vec u) &=& \frac{1}{S_{N-1}}~ \int e^{i\vec{u}\cdot\vec{x}}~ \delta(|\vec x|-R)~ dV
\nonumber \\
&=& \Gamma(N/2)~ (R |\vec u| / 2)^{1-N/2}~ J_{N/2 - 1}(R |\vec u|)\ . 
\end{eqnarray}
Because $\Psi_0$ is the Fourier transform of a non-negative PDF, it automatically satisfies 
Bochner's Theorem, Eq.~(\ref{bochner}).

\section{Infinite Dimensional Limit}
\label{infinite}

We now show that the statistics are Gaussian in the limit of high dimension.  Consider the expansion Eq.~(\ref{seriespsi}).  In the limit $N\rightarrow\infty$, the Taylor series approaches that of a Gaussian:
\begin{equation}
e^{-\frac{u^2}{2}}=\sum_{l=0}^\infty\frac{(-1)^l}{2^{2l}l!}u^{2l}\ .
\end{equation}
Alternatively, we may integrate the PDF over all but the last Cartesian coordinate:
\begin{equation}
P(x_N) = \frac{R~ S_{N-2}}{S_{N-1}}~ \left(1-\left(\frac{x_N}{R}\right)^2\right)^{N-4};\ |x_N| \leq R\ . 
\label{Npdf}
\end{equation}
Since
\begin{equation}
\lim_{N\rightarrow\infty} \left(1-\left(\frac{x_N}{R}\right)^2\right)^{N-4}
= e^{-(N-4) (x_N / R)^2},
\end{equation}
the Gaussian distribution is recovered at large-N.

The fact that a Gaussian distribution appears in the limit $N\rightarrow\infty$ is not unexpected.  Motion along a single coordinate is governed by the essentially random values of the remaining $N - 1$ coordinates.  The Central Limit Theorem then applies, resulting in Gaussian statistics.

\begin{table}
\begin{ruledtabular}
\begin{tabular}{cccc}
Moments & Prediction & Simulation $t_1$ & Simulation $t_2$\\\hline
$\langle x_1 \rangle/R$ & 0 & 0.000610 & 0.0000823\\
$\langle x_1^2 \rangle/R^2$ & $\frac{1}{5}=0.2$ & 0.201 & 0.200\\
$\langle x_1 x_2 \rangle/R^2$ & $0$ & -0.00165 & -0.000211\\
$\langle x_1^3 \rangle/R^3$ & 0 & 0.000505 & 0.0000945\\
$\langle x_1^2 x_2 \rangle/R^3$ & 0 & -0.000209 & -0.0000171\\
$\langle x_1^4 \rangle/R^4$ & $\frac{3}{35}\approx0.0857$ & 0.0866 & 0.0857\\
$\langle x_1^2 x_2^2 \rangle/R^4$ & $\frac{1}{35}\approx0.0286$ & 0.0282 & 0.0286\\
\end{tabular}
\end{ruledtabular}
\caption{A comparison of low order moments, scaled by $R$, as calculated from the exact characteristic functional and from time averages in a direct numerical simulation.  Here $R=17.3$, $t_1=100$ and $t_2=10000$.  Note convergence of the numerical results to the analytical values as the length of the time integration increases.}
\label{moments}
\end{table}

\section{Numerical Analysis}
\label{numerics}

\begin{figure}
\centerline{\includegraphics[scale=0.42]{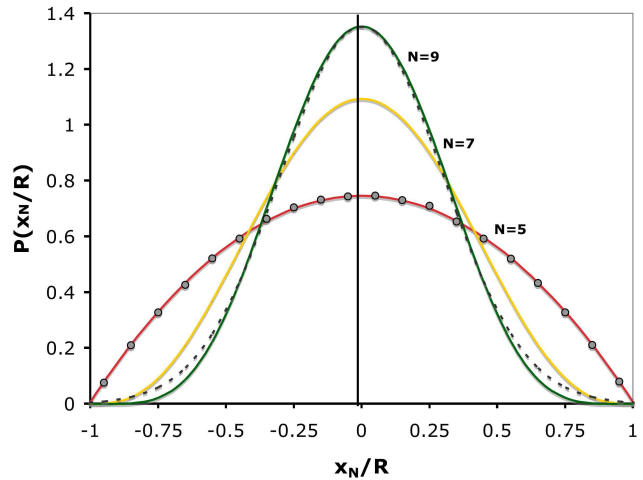}}
\caption{PDF projected onto a single coordinate.  The dots are from direct numerical simulation of the $N = 5$ system; solid lines are plots of Eq.~(\ref{Npdf}) for $N=5,\,7,$ and $9$, showing the approach to Gaussian statistics as $N$ becomes large, and the dashed line is a normalized Gaussian with height equal to the N=9 PDF.}
\label{orszagpdf}
\end{figure}

Finally we make a quantitative comparison with direct numerical simulation.  To be definite consider the $N=5$ case.  The characteristic functional Eq.~(\ref{exactpsi}) in this specific case reduces to: 
\begin{equation}
\Psi_0(\vec u)=3\frac{\sin(R|\vec u|)-R|\vec u|\cos(R|\vec u|)}{R^3|\vec u|^3}
\label{fivepsi}
\end{equation}
or in series form Eq.~(\ref{seriespsi}) to
\begin{equation}
\Psi_0(\vec u)=1-\frac{R^2 |\vec u|^2}{10}+\frac{R^4 |\vec u|^4}{280}-\frac{R^6 |\vec u|^6}{15120} + \ldots
\label{expandedPsi}
\end{equation}
Table~\ref{moments} shows that there is good agreement between moments calculated by successive differentiation of the exact characteristic functional Eq.~(\ref{fivepsi}) and the time-averaged values obtained from direct numerical simulation.  Likewise Eq.~(\ref{Npdf}) gives the projected PDF for the single coordinate $x_5$:
\begin{equation}
P(x_5)=\left\{ \begin{array}{lll} 
\frac{3}{4R} - \frac{3 x_5^2}{4R^3},  &\,& x_5 \leq R\\
0, &\,& x_5 > R \ .
\end{array} \right.
\label{x5pdf}
\end{equation}
Fig.~\ref{orszagpdf} compares this function with the normalized histogram generated from a direct numerical simulation.  Again there is good agreement between the two methods.

\section{Discussion}
\label{discussion}

We employed the Hopf characteristic functional approach to study the statistics of the deterministic Orszag-McLaughlin system.  The system is simple enough, with emergent spherical symmetry in the statistics, that the exact characteristic zero-mode can be found.  Equal-time statistics so obtained are non-Gaussian and are in good agreement with direct numerical simulation.  

Exact solutions serve as rigorous illustrations of physical principles and as benchmarks against which approximate solutions can be checked.  For instance, in the case of the Orszag-McLaughlin system, a cumulant expansion carried out through third order reproduces the exact first through third cumulants (only the second cumulants are non-zero), yet higher-order cumulants are by definition zero.  Inspection of Eq.~(\ref{expandedPsi}) immediately reveals the size of the errors so incurred.  The development of more sophisticated, non-perturbative, methods to address the non-equilibrium statistical mechanics of dynamical systems can likewise benefit from tests against such exact solutions.  

\section{Acknowledgements}
We thank Bernd Braunecker, Matt Hastings, and Peter Weichman for helpful discussions.
This work was supported in part by the National Science Foundation under grant No. DMR-0213818.

\end{document}